\documentclass[twocolumn,showpacs,preprintnumbers,amsmath,amssymb]{revtex4}

\usepackage{graphicx}
\usepackage{dcolumn}
\usepackage{bm}
\usepackage{epsfig}

\begin{document}


\title{A new view of the Lindemann criterion}

\author{U. Buchenau}
 \email{u.buchenau@fz-juelich.de}
 \author{R. Zorn}
\affiliation{%
J\"ulich Center for Neutron Science, Forschungszentrum J\"ulich\\
Postfach 1913, D--52425 J\"ulich, Federal Republic of Germany
}%

\author{M. A. Ramos}
\affiliation{%
Laboratorio de Bajas Temperaturas, Departamento de Fisica de la Materia Condensada, Condensed Matter Physics Center (IFIMAC) and Instituto Nicolas Cabrera, Universidad Autonoma de Madrid, Cantoblanco, E-28049 Madrid, Spain}%

\date{November 19, 2013}

\begin{abstract}
The Lindemann criterion is reformulated in terms of the average shear modulus $G_c$ of the melting crystal, indicating a critical melting shear strain which is necessary to form the many different inherent states of the liquid. In glass formers with covalent bonds, one has to distinguish between soft and hard degrees of freedom to reach agreement. The temperature dependence of the picosecond mean square displacements of liquid and crystal shows that there are two separate contributions to the divergence of the viscosity with decreasing temperature: the anharmonic increase of the shear modulus and a diverging correlation length .
\end{abstract}

\maketitle

According to the hundred years old Lindemann criterion \cite{lindemann}, melting occurs when the thermal motion of the atoms of the crystal reaches a critical mean square displacement of about one tenth of the interatomic distance. It is not very accurately fulfilled \cite{gilvarry,sjoedin}, but it is an intriguing unexplained relation between dynamics and thermodynamics which has always fascinated the physicists working in the field.

Here we reformulate the Lindemann criterion in terms of the average shear modulus $G_c$ of the melting crystal. If one takes the crystal to be a Debye solid and uses the high temperature approximation, the Lindemann criterion reads
\begin{equation}\label{linde}
\langle u_c^2\rangle(T_m)=\frac{3k_BT_m}{M\omega_D^2}\equiv (0.072a)^2,
\end{equation}
where the value 0.072 has been fitted to the data collection of Grimvall and Sj\"odin \cite{sjoedin}, $\langle u_c^2\rangle$ is the mean square displacement in {\it one} direction, the atomic distance $a$ is defined by the atomic volume $v=a^3$, $M$ is the average atomic mass and $\omega_D$ is the Debye frequency.

The Debye frequency is given by the longitudinal sound velocity $v_l$ and the transverse sound velocity $v_t$
\begin{equation}\label{omd}
\omega_D^3=\frac{18\pi^2}{v(1/v_l^3+2/v_t^3)}.
\end{equation}
Taking an average ratio $v_l/v_t$ of 1.8, one gets the mean square displacement
\begin{equation}\label{uq}
	\langle u_c^2\rangle=0.159\frac{k_BT}{G_cv}a^2
\end{equation}
and can express the Lindemann criterion in terms of $Mv_t^2=G_cv$
\begin{equation}\label{lin}
G_cv=31 k_BT_m=51k_BT_g,
\end{equation}
with the approximate relation $T_g=0.6\ T_m$ between melting temperature $T_m$ and glass temperature $T_g$.

The formulation suggests a new view of the Lindemann criterion: it does not indicate an instability of the crystal, but it is rather a necessary condition for the entropy of the liquid. In order to obtain the two stable inherent structures per atom which together with the excess vibrational entropy supply the melting entropy of about $k_B$ per atom, one needs to shear the atomic volumes in the melt by about ten percent. Obviously, constructing a stable solid out of flexible strained units provides much more possibilities than a construction out of rigid units, though one has to pay the price of a considerable strain energy. Above $T_m$, the entropy gain overcompensates this shear energy and makes the liquid the stable thermodynamic state.

\begin{table}[htbp]
	\centering
		\begin{tabular}{|c|c|c|c|c|c|c|c|}
\hline
substance                             & $\rho$  & M   & $T_g$  & $G$ &$\frac{Gv}{k_BT_g}$&$f_s$&$\frac{Gv}{f_sk_BT_g}$\\
\hline
   													          &kg/m$^3$ &  a.u.    &  K     &  GPa   &             &         &              \\	\hline
vit-4$^a$                             & 6112    & 60.0     & 640    &  31.3  &    58.2     &  1      & 58.2         \\
Se                                    & 4167    & 78.96    & 304    &   1.5  &    11.3     &  1/3    & 33.9         \\
SiO$_2$                               & 2198    & 20       &1473    &  35.0  &    26.0     &  5/9    & 46.8         \\
GeO$_2$                               & 3590    & 34.9     & 933    &  21.0  &    26.5     &  5/9    & 47.7         \\
B$_2$O$_3$                            & 1792    & 13.9     & 550    &   5.2  &     8.8     &  1/5    & 44.0         \\
CKN                                   & 2174    & 19.1     & 340    &   4.7  &    14.9     & 19/33   & 25.9         \\
PVC                                   & 1370    & 10.4     & 350    &   1.3  &     3.1     &  1/9    & 27.9         \\
\hline		
		\end{tabular}
	\caption{Measured ratio $Gv/k_BT_g$ in seven glass formers. vit-4 is the metallic glass vitralloy-4, CKN stands for the ionic glass former K$_3$Ca$_2$(NO$_3$)$_7$ and PVC is polyvinylchloride. $^a$ ref. \cite{samwer};  all other data from reference \cite{jcp2012}.}
	\label{tab:Comp}
\end{table}

The glass shear modulus $G$ is usually smaller than $G_c$. Table I lists experimental values of $Gv/k_BT_g$. In the metallic glass vitralloy-4, the value is even higher than the Lindemann prediction of eq. (\ref{lin}), showing that a mixture of atoms with different sizes melts more easily than a pure substance. In a large data collection on metallic glasses \cite{samwer}, the values for $Gv/k_BT_g$ range from 48 to 91. The same tendency is seen in numerical calculations for the Lennard-Jones potential, where one only reaches the glass state with atoms of different sizes \cite{koba,leonforte}, while a pure Lennard-Jones crystal has the same Lindemann mean square displacement at its melting point as pure metals \cite{ljcryst}.

Selenium has a much lower $Gv/k_BT_g$-ratio than the metallic glasses. But selenium has covalent bonds. Each selenium atom is covalently bonded to two neighbors. This implies that one has two hard degrees of freedom per atom (the Se-Se distance and the Se-Se-Se angle) and one soft van-der-Waals degree of freedom; the fraction $f_s$ of soft degrees of freedom per atom is 1/3. In selenium, the frequency of the Se-Se bond stretching vibration is about eight times higher than the van-der-Waals band at low frequencies \cite{sephi}. The Se-Se-Se covalent bending vibration is lower, but still about a factor of three higher than the low frequency band. This implies that 90 \% of the mean square displacement is due to the van-der-Waals bonds. One degree of freedom supplies virtually the whole mean square displacement and eq. (\ref{uq}) is derived for three equivalent degrees of freedom per atom. Consequently, $Gv/k_BT_g$ has to be corrected by a factor of 3. Indeed, $Gv/f_sk_BT_g=33.9$ is about two thirds of the Lindemann prediction for the crystalline shear modulus.

The same argument holds for other covalently bonded glass formers. In the next two examples, silica and germania, the four Si-O stretching degrees of freedom are markedly harder than the five O-Si-O bending vibrations \cite{jcphil,thorpe}. In B$_2$O$_3$, one excludes the B-O stretching as well as the O-B-O bending. This leaves three van-der-Waals degrees of freedom per B$_2$O$_3$-unit, arriving at $f_s=1/5$.  For the ionic glass former CKN, one considers the NO$_3$ unit as a rigid molecule with only six degrees of freedom, reducing the number of degrees of freedom by 19/33. The polyvinylchloride monomer C$_2$H$_3$Cl has only the two C-C rotations as soft degrees of freedom, which implies $f_s=1/9$, again in reasonable agreement with the Lindemann criterion. The average value for the seven glass formers in Table I is $Gv/f_s=41\ k_BT_g$, about 80 \% of the crystalline value of eq. (\ref{lin}).

It is tempting to look for a connection between this value and the effective energy barrier for the flow at $T_g$, which happens to have a value nearby, about 36 $k_BT_g$. But in the Lindemann relation $Gv$ is a force constant, not a barrier. Converting it into a barrier requires a microscopic consideration:

\begin{figure}[b]
\hspace{-0cm} \vspace{0cm} \epsfig{file=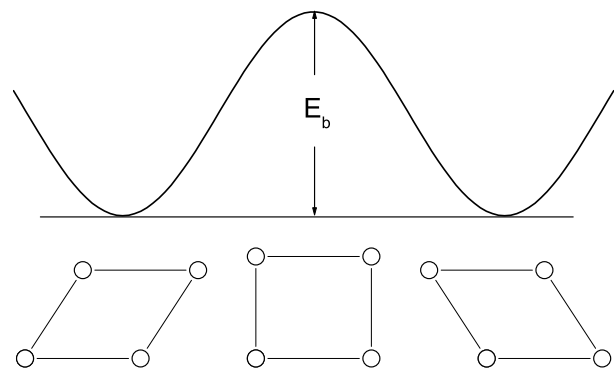,width=8 cm,angle=0} \vspace{0cm} \caption{The potential energy of a group of four atoms in close packing as a function of the local shear.}
\end{figure}

Consider the four neighboring atoms shown in Fig. 1, undergoing a shear transformation from the stable configuration on the left to another stable configuration on the right. The second derivative of their potential in the shear angle is $4Gv$. The difference between the shear angle of the two stable configuration is 60 degrees, in radian units close to 1. For a cosine potential, the corresponding barrier height $E_4$ is 2$Gv/\pi^2$.

Of course, such a structural jump is not possible within a stable solid. In fact, the energy maximum configuration of the square in the middle is even stable in an fcc crystal, because the elastic matrix around the square compensates the negative spring inside. The same is true for the gliding triangle motion of six atoms \cite{buscho}, which converts an octahedron into a bitetrahedron. In this case, the barrier is only 3$Gv/2\pi^2$ (6 $k_BT_g$ for $Gv=41\ k_BT_g$), because the shear angle from one stable configuration to the other is smaller.

While a single four-atom or six-atom jump is not possible, it seems likely that the real structural transitions at $T_g$ are combinations of several such jumps within a central core, leading to a new core which again fits reasonably well into the surrounding elastic matrix. To obtain the total barrier $E_b$ of about 36 $k_BT_g$ needed to inflate the microscopic time constant $\tau_0=10^{-13}$ seconds to the relaxation time of hundred or thousand seconds at the glass transition, one has to postulate a combination of about six elementary six-atom jumps, consistent with an inner core of twenty to forty atoms.

In this picture, the energy barrier for the flow is not only proportional \cite{dyre} to the temperature-dependent modulus $G$, but also to the number $n_s$ of four-atom or six-atom jumps within the central core (the cooperatively rearranging region \cite{dyna}), which might increase with decreasing temperature. An increase and even a divergence of a dynamic correlation length with decreasing temperature has been first postulated by Adam and Gibbs \cite{dyna}. It is a tempting idea: the structural entropy extrapolates to zero at the Kauzmann temperature, so one has less and less possibilities to jump into another structural state. Then the structural reorganization requires larger and larger cooperatively rearranging volumes. The Adam-Gibbs concept is supported by numerical calculations, which have been able to see the increase of the correlation length with decreasing temperature in various ways \cite{sastry,biroli}.

Following Jeppe Dyre \cite{dyre}, we define the fragility in terms of the negative logarithmic derivative $I$ of the flow barrier with respect to temperature ($I=(m-16)/16$ in terms of the usual measure $m$ of the fragility \cite{bohmer}). The flow barrier (taking six-atom units)
\begin{equation}\label{eb}
	E_b=\frac{3n_s}{2\pi^2}Gv
\end{equation}
is proportional both to $n_s$ and to $T/\langle u^2\rangle$ (via $Gv$ and eq. (\ref{uq})). Therefore it has the negative logarithmic derivative
\begin{equation}\label{ebln}	-\frac{\partial\ln{E_b}}{\partial\ln{T}}=-\frac{\partial\ln{n_s}}{\partial\ln{T}}+\left(\frac{\partial\ln{\langle u^2\rangle}}{\partial\ln{T}}-1\right)
\end{equation}

\begin{figure}[b]
\hspace{-0cm} \vspace{0cm} \epsfig{file=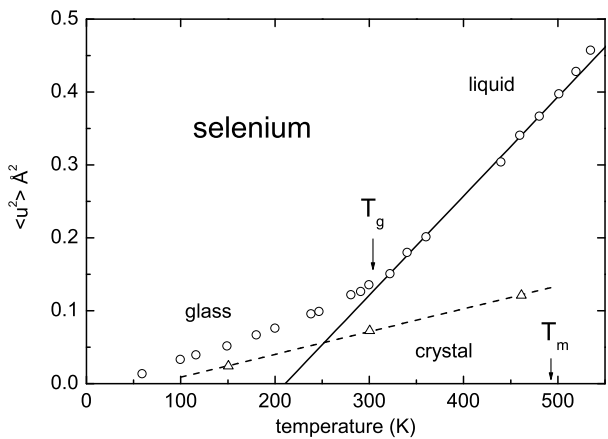,width=8 cm,angle=0} \vspace{0cm} \caption{The mean square displacements in crystalline, glassy and liquid selenium \cite{zorn} Note that the liquid data extrapolate to the crystalline ones at 252 K.}
\end{figure}

\begin{figure}[b]
\hspace{-0cm} \vspace{0cm} \epsfig{file=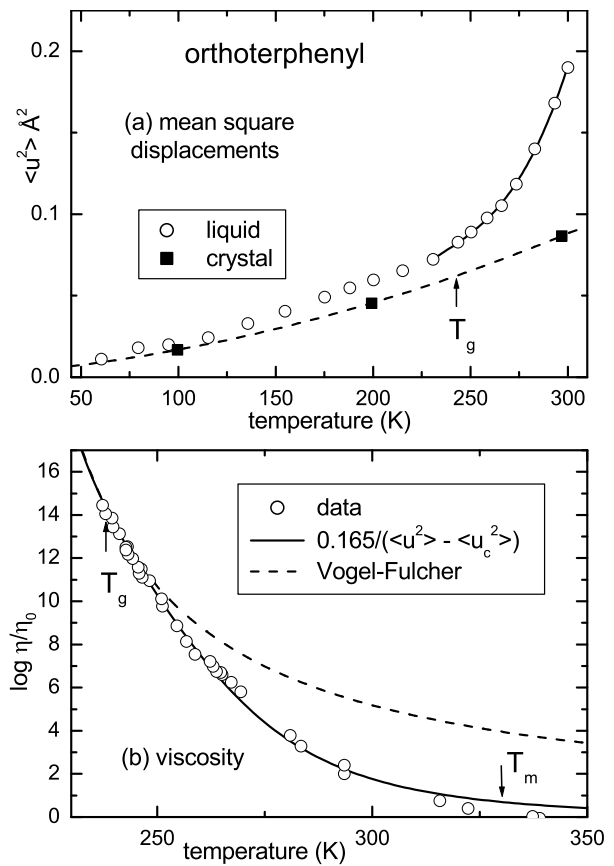,width=8 cm,angle=0} \vspace{0cm} \caption{(a) The mean square displacements in crystalline, glassy and liquid ortoterphenyl \cite{toelle}. The lines are the corresponding fits. (b) Proportionality of the logarithm of the viscosity \cite{plazek} $\eta/\eta_0$ with $\eta_0=0.2\ Pa\ s$ to the inverse difference between liquid and crystalline mean square displacements (the continuous line). The dashed line is the Vogel-Fulcher relation obtained by linearizing the mean square displacements around $T_g$.}
\end{figure}

At first sight, this separation in two contributions seems to contradict the experimental finding of a proportionality of the flow barrier to the shear modulus alone in many substances \cite{dyre}. This discrepancy, however, has been resolved by a recent thorough investigation of mean square displacements at different energy resolution \cite{niss}. One needs a resolution corresponding to a time scale of at least several nanoseconds to see the mean square displacement of the macroscopic shear modulus. The temperature dependence of the quantity $\langle u^2 \rangle/T$ on the picosecond level is always too weak to explain the full fragility. This has also been seen in a recent data collection \cite{dino}, which postulated a proportionality of the logarithm of the viscosity to $\alpha+\beta/\langle u^2 \rangle+\gamma/\langle u^2 \rangle^2$. The term $\gamma/\langle u^2 \rangle^2$ introduces a similar effect as the temperature dependence of $n_s$ and dominates the behavior at $T_g$.

The best experiment for a check of these ideas is a twenty-year old determination \cite{zorn} of the mean square displacements in glassy, liquid and crystalline selenium (Fig. 2). The logarithmic derivative of the liquid $\langle u^2\rangle$ at $T_g=304$ K is 3.1, explaining 2.1 units of the total fragility $I=4.4$ (the usual fragility \cite{bohmer} m is 87). Thus more than half of the fragility remains unexplained and must be attributed to a logarithmic decrease of $n_s$ of 2.3 with temperature (note that this information on the volume of the cooperatively rearranging region comes directly from the pair correlation function; it is not necessary to invoke higher correlations \cite{cecile,fragiadakis}). 

The mean square displacement of the undercooled liquid extrapolates to the one of the crystal at 252 K, close to the Kauzmann temperature of 240 K where the excess entropy over the crystal extrapolates to zero and the Vogel-Fulcher temperature of 251 K where the viscosity extrapolates to infinity \cite{angell}. In fact, one finds a proportionality of the logarithm of the viscosity ratio $\eta/\eta_0$ ($\eta_0$ high temperature viscosity, which in selenium is 3.1$\cdot 10^{-4}\ Pa\ s$) to the inverse difference between liquid and crystalline mean square displacements \cite{zorn}. It holds over eighteen decades of viscosity variation, from the aging regime below $T_g$ up to a temperature high above the melting temperature, providing a much better fit of the viscosity than any Vogel-Fulcher law. 

The same proportionality between the logarithm of the viscosity ratio \cite{plazek} and the inverse difference between liquid and crystal mean square displacement is found in orthoterphenyl (OTP). The mean square displacements \cite{toelle} in Fig. 3 (a) show a much stronger curvature at $T_g$ than those of selenium. Nevertheless, if one fits $\langle u^2\rangle$ in terms of a third order function (the continuous line in Fig. 3 (a)) and $\langle u_c^2\rangle$ in terms of a second order function in temperature (the dashed line in Fig. 3(a)), one finds again the proportionality shown by the continuous line in Fig. 3 (b). The fit with $\eta_0=0.2\ Pa\ s$ fails above $T_m$, but still covers twelve decades of viscosity variation.

One obtains a clearer understanding of these two experimental findings assuming that $n_s$ is proportional to the inverse difference between crystal and liquid shear modulus $G_c-G$. In the Adam-Gibbs reasoning \cite{dyna}, $n_s$ is inversely proportional to the structural entropy difference of liquid and crystal. A proportionality of the shear modulus difference to the entropy difference sounds plausible. The assumption provides the extrapolated divergence of $n_s$ at $G_c=G$ which one obviously needs to understand the strong tendency to a divergence of the viscosity close to $T_g$. Since $E_b$ is proportional to the product of $n_s$ and $Gv$
\begin{equation}\label{fin}
	\log{\eta/\eta_0}\propto \frac{E_b}{T}\propto\frac{Gv}{T(G_c-G)}\propto\frac{1}{\langle u^2\rangle-\langle u_c^2\rangle},
\end{equation}
if one neglects the weak temperature dependence of the crystal shear modulus $G_c$. Note that a linearization of both $\langle u^2\rangle$ and $\langle u_c^2\rangle$ around $T_g$ converts this relation into the Vogel-Fulcher law $\log{\eta/\eta_0}\propto 1/(T-T_0)$, thus identifying the Vogel-Fulcher temperature $T_0$ with the point where the extrapolated mean square displacement of the liquid reaches the crystalline one.

To summarize, the Lindemann criterion has its physical basis in the entropy of the liquid: a temperature able to reach a mean square displacement of one tenth of the interatomic distance in the crystal is also able to distort the structural units of the liquid by a shear angle of about one tenth. The shear flexibility allows to form a much larger number of stable inherent structures in the liquid than those accessible to rigid units, enhancing the liquid entropy to the point where it compensates the shear energy and makes the liquid the stable phase.

One can use the Lindemann criterion to estimate the flow barrier in the liquid. The estimate requires a geometrical consideration of the elementary relaxing units, of which several have to combine to form a cooperatively rearranging region. On the basis of this concept, one finds a physical explanation for the proportionality of the logarithm of the viscosity ratio $\eta/\eta_0$ ($\eta_0$ high temperature viscosity) to the inverse of the difference of the picosecond mean square displacements of crystal and liquid. A measurement of these two quantities allows to determine the volume of the cooperatively rearranging region. The heavily discussed fragility of undercooled liquids is due to a combination of two classical mechanisms, the Adam-Gibbs mechanism of a diverging correlation length and the proportionality of the flow barrier to the picosecond shear modulus.

\end{document}